\documentclass{elsart}

\usepackage{natbib}

 \usepackage{graphicx}

\usepackage{amssymb}

\begin{document}

\begin{frontmatter}



\title{Nested structure acquired through simple evolutionary process}


\author[Tokyo,corr]{Kazuhiro Takemoto}
\ead{takemoto@cb.k.u-tokyo.ac.jp}
\author[Tokyo,PSC,Keio]{Masanori Arita}


\address[Tokyo]{Department of Computational Biology, Graduate School of Frontier Sciences, University of Tokyo, Kashiwanoha 5-1-5, Kashiwa, Chiba 277-8561, Japan}

\address[PSC]{RIKEN Plant Science Center, Suehiro-cho 1-7-22, Tsurumi-ku, Yokohama, Kanagawa 230-0045, Japan}

\address[Keio]{Institute for Advanced Biosciences, Keio University, Baba-cho 14-1, Tsuruoka, Yamagata 997-0035, Japan}

\corauth[corr]{Corresponding author.}

\begin{abstract}
Nested structure, which is non-random, controls cooperation dynamics and biodiversity in plant-animal mutualistic networks.
This structural pattern has been explained in a static (non-growth) network models.
However, evolutionary processes might also influence the formation of such a structural pattern.
We thereby propose an evolving network model for plant-animal interactions and show that non-random patterns such as nested structure and heterogeneous connectivity are both qualitatively and quantitatively predicted through simple evolutionary processes.
This finding implies that network models can be simplified by considering evolutionary processes, and also that another explanation exists for the emergence of non-random patterns and might provide more comprehensible insights into the formation of plant-animal mutualistic networks from the evolutionary perspective.
\end{abstract}


\end{frontmatter}

\section{Introduction}
Elucidation of basic design principles behind mutualistic networks is a central topic in theoretical ecology for understanding cooperation dynamics and biodiversity.
In particular, {\it nested structure} is a well-known non-random structural pattern observed in plant-animal mutualistic networks \citep{Bascompte2003}.
In this structure, pollinators (animals) of a certain plant form a subset of those of another plant in a hierarchical fashion.
Such non-random patterns often strongly control dynamics of ecological systems. 
For example, nested structure is believed to minimize competition and increase biodiversity \citep{Bastolla2009}.
In addition to nestedness, the heterogeneous connectivity, which is defined as the occurence of diversified patterns of interaction among species (both plants and animals), is found in such mutualistic networks \citep{Jordano2003,Saavedra2009}.
Together with nestedness, heterogeneous connectivity might influence the dynamics of ecological systems.

To elucidate the origin of non-random patterns such as nested structure and heterogeneous connectivity, \citet{Saavedra2009} recently proposed the bipartite cooperation (BC) model inspired by food-web models based on species traits and external factors (reviewed in \citet{Stouffer2005}).
Although this model agrees well with real plant-animal mutualistic networks, it is a static (non-growth) network model in which the number of species (plants and pollinators) is fixed.
The structure of model-generated networks is determined by three types of parameters drawn from exponential or beta distributions: foraging traits (e.g., efficiency and morphology), reward traits (e.g., quantity and quality) and external factors such as environmental context (e.g., geographic and temporal variation).

The structure of interactions between plants and animals may be affected by not only species traits and external factors but also evolutionary history \citep{Rezende2007-1}.
Thus far, few efforts have been made to theoretically explain the development of interactions between plants and animals.
For example, \citet{Rezende2007-2} showed that plant-animal mutualistic networks are qualitatively more nested by using phylogenetic information in addition to species traits.
Futhermore, they repoted that the phylogenetic signal was reported to be significant but weak in plant-pollinator networks.
In addition, by using evolving network models, \citet{Guimaraes2007} explained the origin of frequency distributions of the number of interactions per species following a power law with an exponential truncation (i.e., heterogeneous connectivity).
However, this previous model was less predictive than the BC model, is a static (non-growth) network model \citep{Saavedra2009}.
Therefore, these results suggest that the evolutionary effects in plant-animal mutualistic networks are not accurate.

Nevertheless, evolutionary history may also determine the structure of plant-animal mutualistic networks because species are believed to mutually interact via traits or functions acquired through evolution.
Ecological systems are very complex, and we have not yet understood the whole picture of ecological systems.
To understand such complex systems, we need to consider them from several perspectives.
Therefore, it is important to develop good evolving network models that consider the effect of evolutionary processes on network formation.
In the present study, we present an evolving network model of plant-animal interactions that takes into account simple evolutionary processes and use this model to investigate the effect of evolutionary history on the formation of plant-animal mutualistic networks.
Numerical simulations showed that this simple evolving network model better reproduced nestedness and heterogeneous connectivity by using real data on plant-animal mutualistic networks.
This result implies a not negligible contribution of evolutionary history to the structure of networks.
Furthermore, we discuss the comprehensible origin of nested structures and heterogeneous connectivity from an evolutionary perspective.

\section{The Model}
\begin{figure}[tbhp]
\begin{center}
	\includegraphics{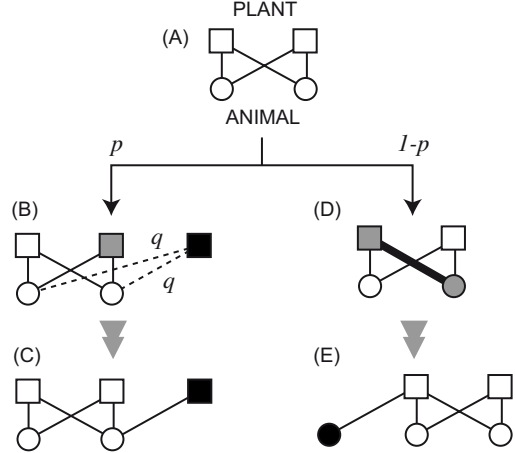}  
	\caption{
	Schematic diagram of our model. The squares and circles represent plants and animals, respectively.
	(A) Initial plant-animal network is represented as a complete bipartite network.
	(B and C) The addition of a new plant and the resulting plant-animal interactions.
	The gray square represents a randomly selected plant.
	The filled square indicates a new plant resulting from duplication of the selected plant.
	The dashed lines represent possible interactions between the new plant and animals.
	(D and E) The addition of a new animal and the resultant interaction.
	The thick link between the gray nodes corresponds to a randomly selected plant-animal pair. The filled circle indicates a new animal.
	}
	\label{fig:model}
\end{center}
\end{figure}

We developed an evolving network model based on the model proposed by \citet{Takemoto2009} to explain plant evolution.
In our model, a small initial plant-animal mutualistic network (Fig. \ref{fig:model} A) is preparated and represented as a complete bipartite graph with $n_0$ plants and $n_0$ animals, where $n_0$ is the initial number of species, and is an integer $\geq 1$.
This initial network evolves according to two simple evolutionary mechanisms as follows:

(i) Animals interacting with new plants are inherited from those interacting with existing (ancestral) plants.
This mechanism is based on the phylogenetic resemblance between related species \citep{Garland2005}.
We assume that new plants emerge because of mutations in ancestral plants.
In our model, mutations in plants occur at the probability $p$ and time $t$, and new plants originate from randomly selected existing plants.
In this case, the traits (e.g., forms, colors, and chemical compounds) of new plants are similar to those of ancestral plants because of the phylogenetic resemblance.
Since plant-animal interactions are based on these traits, the animals connecting to the new species are inherited from those connecting to ancestral plants (Fig. \ref{fig:model} B).
By considering divergence, however, we model each animal such that it is
inherited from animals interacting with the ancestral plant at the probability $q$ (Fig. \ref{fig:model} C).
This indicates that plants lose some traits due to the divergence.
As explained above, since plants interact to animals via their traits, the plants also lose some interactions with animals. 
In addition, a bipartite network model generated based on the above inheritance (or copy) mechanism was proposed in \citet{Nacher2009} to describe the evolution of protein domain networks at around the same time that we developed our model.

(ii) Plants acquire new interacting animals after obtaining new traits (phenotypes) because plants are believed to connect to animals via phenotypic information as above.
In particular, chemical compounds, which are often flower pigments, are involved in attracting insects (animals), protecting plants from insects \citep{Harbone2000}.
During evolution, living organisms (plants) are believed to obtain new phenotypes (traits) via genetic changes such as gene duplications and horizontal gene transfers.
Since new phenotypes (traits) are based on existing phenotypes, plants interact with new animals.
We modeled the above event to occur at the probability $1-p$ and time $t$; a plant-animal pair is selected at random (Fig. \ref{fig:model} D) because new traits are based on existing traits, and thus, a plant acquires a new animal (Fig. \ref{fig:model} E).

Our model has two parameters $p$ and $q$. We can generate the model network by estimating these parameters using observable parameters in real plant-animal mutualistic networks: the number of plants $P$, the number of animals $A$, and the number of interactions $L$.

Parameter $p$ is estimated as
\begin{equation}
p=\frac{P}{P+A}
\label{eq:p}
\end{equation}
where $P=pt$ and $A=(1-p)t$ in our model.

To estimate parameter $q$, we need to consider the time evolution of $L$. This
is derived as $L\approx (1-p)t/(1-q)$ \citep{Takemoto2009}. Since $A=(1-p)t$, parameter $q$ is estimated as
\begin{equation}
q=1-\frac{A}{L}
\label{eq:q}
\end{equation}
Using Eqs. (\ref{eq:p}) and (\ref{eq:q}), we estimated parameters $p$ and $q$ from real data and
generated corresponding model networks for comparison with real ones.



\section{Results and Discussion}
We first show that our model-generated networks reproduce nested structures by using real data on plant-animal mutualistic networks reported by \citet{Saavedra2009}.
The nested structure indicates that animals (pollinator) of a certain plant form a subset of those of another plant.
The degree of nested structure is measured by nestedness $N$, ranging from perfect non-nestedness ($N=0$) to perfect nestedness ($N=1$).
When $N$ is close to $1$, we can see such a strong containment relationship in a hierarchical fashion.
When $N$ is close to $0$, in contrast, plants have independent sets of animals: there is no such a relationship.
In order to measure $N$ of model and real-world networks, we used the BINMATNEST program \citep{Girones2006}.

We also assessed the BC model and random networks generated using the null model II \citep{Bascompte2003} for comparison.
The null model generates random bipartite networks with degree sequences that are similar to those of real networks.
In the null model, the probability that a plant connects to an animal is proportional to the product of node degrees for a given plant and animal.

\begin{figure}[tbhp]
\begin{center}
	\includegraphics{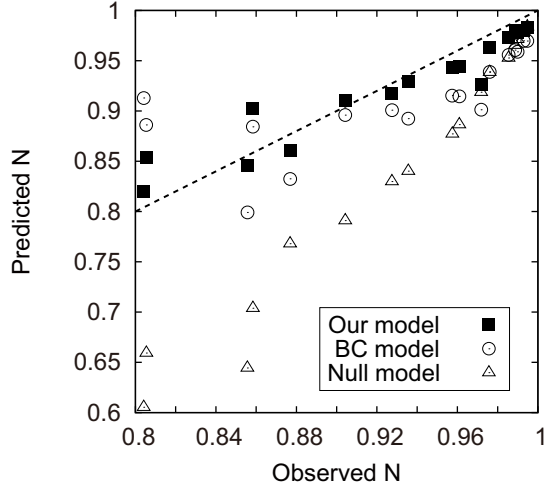}  
	\caption{
	Comparison of nestedness between models and real data.
	The dashed line represents the perfect agreement between predicted nestedness and observed nestedness.
	The values of nestedness from models are averaged over 100 realizations.
	}
	\label{fig:nestedness}
\end{center}
\end{figure}

The comparison of nestedness between models and real data is shown in Fig. \ref{fig:nestedness}.
Our model and the BC model produced comparable nestedness in real network data.
The null model could not predict nestedness.
This result indicates that nested structure is independent from degree sequences and that the predictions of our model and the BC model are nontrivial.

Furthermore, we closely evaluated the prediction accuracy of our model and the BC model by using the Pearson correlation coefficient and root mean square error between predicted nestedness $x_i$ and observed nestedness $y_i$, defined as
\[
\mathrm{RMSE}=\sqrt{\frac{1}{n}\sum_{i=1}^n(x_i-y_i)^2},
\]
where $n$ is the number of samples.
As shown in Table \ref{table:compari_nest}, our model had higher prediction accuracy than the BC model. 

\begin{table}[tbhp]
\caption{Comparison of prediction accuracy between our model and the BC model.}
\label{table:compari_nest}
\begin{center}
\begin{tabular}{lcc}
\hline
\hline
& Correlation Coefficient & Root Mean Square Error \\
\hline
Our model & {\bf 0.934} & {\bf 0.0232} \\
BC model & 0.790 & 0.0415 \\
\hline
\hline
\end{tabular}
\end{center}
\end{table}

Our model can clearly explain the origin of nested structure in plant-animal mutualistic networks by using fewer parameters.
In this model, animals interacting with a new plant are inherited from those interacting with an ancestral plant because the two types of plants tend to be similar.
New plants acquire some of the interacting animals via divergence (elimination of interactions).
As a result, animals of an offspring plant become a subset of those of the parent plant, and thereby resulting in nestedness.



\begin{figure*}[tbhp]
\begin{center}
	\includegraphics{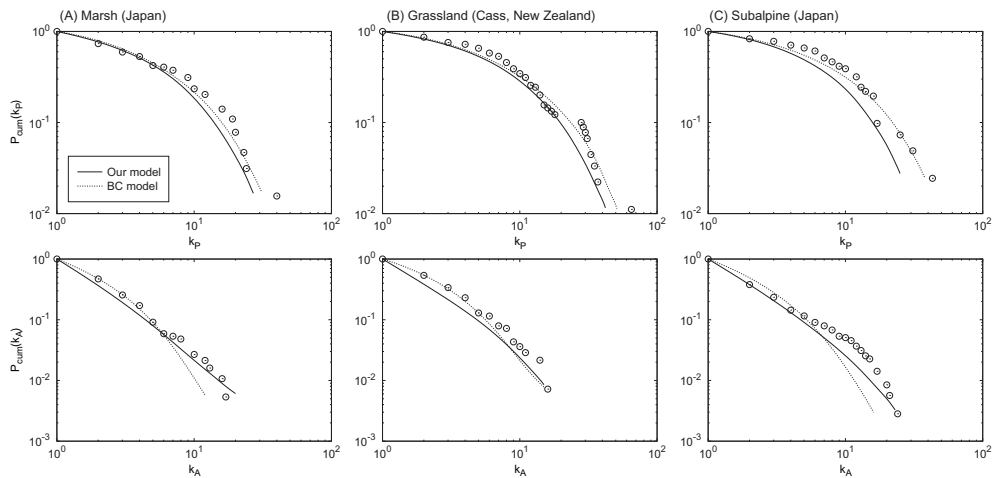}  
	\caption{
	Cumulative degree distributions in plant-animal mutualistic networks for plants (top column) and pollinators (animals) (bottom column).
	The circles and lines represent real data and models, respectively.
	The cumulative distributions of models are averaged over 100 realizations.	}
	\label{fig:degree}
\end{center}
\end{figure*}

Next, we investigated the frequency distribution of the number of interactions per species (degree distribution).
In plant-animal mutualistic networks, we often observe heterogeneous degree distributions, which follow a power law with an exponential truncation.
Our model also reproduces heterogeneous distributions \citep{Takemoto2009}, which are not influenced by the structure of initial networks (i.e., $n_0$) in the case of large $P$ and large $A$.  
The cumulative degree distributions of real plant-animal mutualistic networks (symbols) and model-generated networks (lines) are shown in Fig. \ref{fig:degree}.
The cumulative distributions of only three representative networks are shown owing to space limitations.
We could observe cumulative distributions of two types [$P_{\mathrm{cum}}(k_P)$ and $P_{\mathrm{cum}}(k_A)$, where $k_P$ and $k_A$ denote the degrees on nodes corresponding to plants and animals (pollinators), respectively] because plant-animal mutualistic networks are represented as bipartite graphs.
The models studied were in good agreement with real data (Fig. \ref{fig:degree}).

In order to quantitatively verify goodness of fits between models and real data, we calculated the Kolmogorov-Smirnov (KS) statistics (distance) between empirical distributions and predicted distributions for $P_{\mathrm{cum}}(k_P)$ (KS$_P$) and $P_{\mathrm{cum}}(k_A)$ (KS$_A$).
We also investigated the relationship between network size and the KS distance of degree distributions.
Here, the network size and prediction accuracy were defined as $P+A$ and the sum of the KS distances for the two distributions (i.e. $\mathrm{KS}_P+\mathrm{KS}_A$), respectively.

\begin{figure}[thbp]
\begin{center}
	\includegraphics{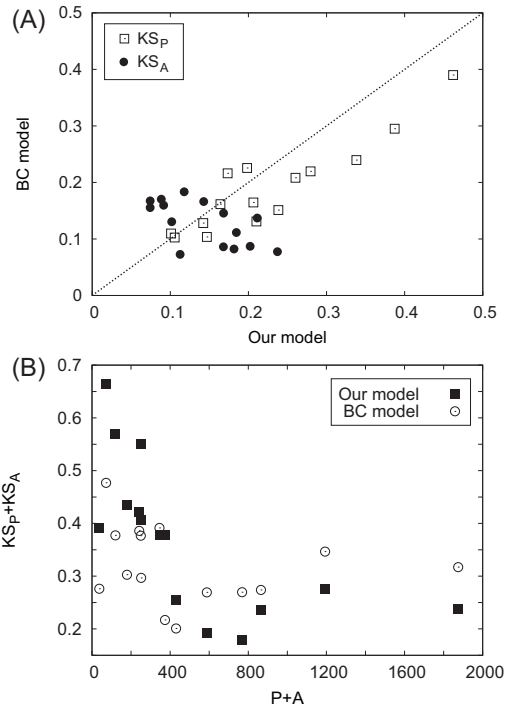}  
	\caption{
	Kolmogorov-Smirnov (KS) statistics (distances) between models and real data.
	(A) Comparison of KS distance between our model and the BC model.
	The dashed line indicates that our model and the BC model had the same prediction accuracy.
	KS$_P$ and KS$_A$ are KS distances between observed and predicted distributions for $P_\mathrm{cum}(k_P)$ and $P_\mathrm{cum}(k_A)$, respectively.
	(B) Correlation between network size and KS distance.
	}
	\label{fig:KS_prob}
\end{center}
\end{figure}

The KS distances in our model and the BC model are compared in Fig. \ref{fig:KS_prob} A.
Overall, the KS distances were small for both models, i.e., in good agreement with real data.
However, we also observed some low-accuracy predictions in both cases for small-sized networks.
The KS distance increased as the network size decreased (Fig. \ref{fig:KS_prob} B) because the structure of small model-generated networks may have strong fluctuated owing to its stochastic nature.
The accuracy of our model was lower than that of the BC model for small networks.
This might be because the BC model considers more parameters such as species traits and external factors than our model dose and is more independent from initial networks.
In the case of large networks, however, our model more precisely reproduced the degree distributions than did the BC model.

Using our model, we could elucidate the origin of heterogenous distributions, which generally corresponds to multiplicative processes or preferential attachments \citep{Barabasi2004}, in plant-animal mutualistic networks from an evolutionary perspective.
When new animals emerge, plants establish new interactions with them.
Plants that interact many animals are likely to acquire more new animals because plant-animal pairs are selected at random.
This phenomenon reflects preferential attachment, which is similar to that in the Dorogovtsev-Mendes-Samukhin model \citep{Dorogovtsev2001}.
A similar mechanism exists in the case of animals.
In our model, new plants inherit the animals of randomly selected plants.
That is, animals also established new interaction with plants when new plants emerge because of random mutation.
Thus, animals that interact with many plants tend to acquire more new plants.
This is similar to preferential attachment in the duplication-divergence model \citep{Vazquez2003}.

Our model is simpler than the BC model.
In the BC model, parameters such as species traits and external factors have to be expressly provided; these parameters need to drawn from given probabilistic distributions.
Furthermore, the interactions between plants and animals are determined according to the linking mechanisms based on such parameters.
In contrast, species traits and external factors and linking mechanisms are masked to a certain extent in our model, which takes into account evolutionary processes.
Therefore, the above parameters, probabilistic distributions, and linking mechanisms based on parameters need not be directly considered in our model.
However, this implies that species traits and external factors are {\it not} inessential.
Species traits and external factors are important for plant-animal interactions.
In our model, plant-animal interactions are determined as a result of species traits and external factors.
However, our model has no parameters corresponding to these traits and factors.
Thus, it does not reveal the relationship of species traits and external factors with network structure in exchange for the simplification.
This is a limitation of our model.

We based our model on plant evolution. 
As explained in Sec. 2, plant chemical compounds are critical for plant-animal interactions \citep{Harbone2000} and hence, such interactions are often explained from the perspective of plants.
Thus, we assumed that interactions between plants and animals are influenced by the traits of plants.
Nonetheless, network formation may also be influenced by animal evolution.
In this case, however, a more complicated model is required (i.e., more parameters).
Therefore, we selected a model based on plant evolution.
Due to the simplification, our model does not clearly explain animals.
For example, it does not consider what kind of animals join in networks when new species emerge.
In other wards, it does not distinguish whether the new animals are related to existing ones or not, and it is not clear what traits animals have.
For this reason, the strength of phylogenetic signal is not evaluated.
Therefore, traits of animals and phylogenetic relationships in plant-animal mutualistic networks are not discussed by using our model.
This is also a limitation of our model, and it is a problem needing to be solved in the future.

It is surprising that the same model can explain two different systems: plant-animal mutualistic networks and species-metabolite relationship in regards to flavonoids \citep{Takemoto2009}.
A hidden evolutionary relationship between flavonoids and animals, through a radical speculation, may exist.
Some flavonoids, such as flower pigments, play roles in attracting insects (pollinators) and protecting plants from insects.
Such chemical compounds may correspond to traits (or phenotypes) of plants.
Thus, the structure of species-metabolite relationships is expected be similar to that of plant-animal networks.

As shown above, structural patterns are also reproduced through simple evolutionary processes.
However, \citet{Rezende2007-2} showed that the phylogenetic signal was significant but weak in an observed plant-animal mutualistic network.
Our model, which shows high phylogenetic signal, does not explain this network well.
This may be cause that there are several independent mechanisms for generating nested structure and heterogeneous connectivity.
Network formation may be explained several independent theories.
One is described by the BC model, which implies low phylogenetic signal.
The other is based on our model, which suggests high phylogenetic signal.
In nature system, the nested structure and heterogeneous connectivity may be independently generated by both mechanisms.
Due to the balance of mechanisms, thus, we can expect to observe plant-animal mutualistic networks with high phylogenetic signal in addition to those with low phylogenetic signal.
We believe that there are other plant-animal mutualistic networks descried by our model.
Species traits, which are important for plant-animal interactions, are inherited from ancestral species to new species. 
Thus, evolutionary history may greatly affect the formation of plant-animal mutualistic networks.
However, earlier models do not reveal the effect of evolutionary history on ecological networks because they do not clearly consider evolution.
We believe that our model (an evolving network model) is important to obtain a deeper understanding of the formation of plant-animal mutualistic networks.


Regarding external factors, our model has a limitation.
Heterogeneous environments (e.g., geographic, temperature, and nutrient differences) may reduce phylogenetic signals in plant-animal mutualistic networks.
Our model may not clearly explain the formation of networks under such situations because it has no parameter corresponding to external factors due to the simplification.
Our model can be applied to ecological networks if plant-animal relationships are comprehensively obtained under ideal conditions (e.g. environmentally homogeneous islands).



For simplicity, we did not consider a number of important evolutionary processes such as extinction and convergent evolution.
In particular, degree distributions might be altered because of such extinctions \citep{Enemark2007}.
However, our results indicate that such mechanisms might have only negligible effects on the structural properties of ecological networks such as nested structure and heterogeneous connectivity.
In plant species, genome doubling (polyploidity) is a major driving force for increasing genome size and the number of genes \citep{Adams2005}.
Duplicated genes typically diversify in function, and some are phenotypically expressed as forms, colors, syntheses of chemical compounds, etc.
Such phenotypes (traits), especially the syntheses of chemical compounds, are therefore expected to increase, indicating that the effect of node losses may be disregarded when considering the global tendencies of plant-animal mutualistic networks.
However, this implies that deletions of nodes and interactions are {\it not} unnecessary.
Such evolutionary mechanisms might play important roles in determining partial (or local) interaction patterns of bipartite relationships.
Thus, we need to focus on such evolution processes in the future to fully understand the formation of plant-animal mutualistic networks.

Although our model has the above limitations due to simplification of model, it may be extensible by considering several factors (parameters).
We believe that our model helps to reveal the relationships such factors and evolutionary history in plant-animal mutualistic networks and to more deeply understand the formation of networks in the future.

\section{Conclusion}
By considering simple evolutionary processes, we have shown that non-random structural properties of plant-animal mutualistic networks, such as nested structures and heterogeneous connectivity, can be both qualitatively and quantitatively predicted.
This finding suggests that network models can be simplified by considering evolutionary processes. 
Furthermore, it implies that another explanation exists for the emergence of non-random structures (i.e., from the evolutionary perspective).
\citet{Rezende2007-1,Rezende2007-2} and our simple description of interactions between plants and animals indicate the importance of evolutionary history, which is worth further investigation. 
The evolving network model does not contradict static (non-growth) network models because these models describe respective different mechanisms for formation of non-random structures, and therefore provides additional insights into the formation of plant-animal mutualistic networks.

\section*{Acknowledgments}
We thank S. Saavedra for providing data on plant-animal networks and helpful comments.
We are also obliged to the reviewers for their valuable discussion and comments.

\end{document}